\documentclass[twocolumn,showpacs,preprintnumbers,amsmath,amssymb]{revtex4}

\usepackage[pdftex]{graphicx}
\usepackage{dcolumn}
\usepackage{bm}

\begin{document}

\preprint{APS/123-QED}

\title{Superlinear Scaling for Innovation in Cities}

\author{Samuel Arbesman}
\email{arbesman@hcp.med.harvard.edu}
 \affiliation{Department of Health Care Policy, Harvard Medical School, 180 Longwood Avenue, Boston, Massachusetts 02115}
\author{Jon M. Kleinberg}%
\email{kleinber@cs.cornell.edu}
\affiliation{Computer Science, Cornell University, Ithaca, New York 14853}

\author{Steven H. Strogatz}
 \email{shs7@cornell.edu}
\affiliation{Theoretical and Applied Mechanics, Cornell University, Ithaca, New York 14853}

\date{\today}

\begin{abstract}
Superlinear scaling in cities, which appears in sociological quantities such as economic productivity and creative output relative to urban population size, has been observed but not been given a satisfactory theoretical explanation. Here we provide a network model for the superlinear relationship between population size and innovation found in cities, with a reasonable range for the exponent.
\end{abstract}

\maketitle

\section{Introduction}

It has been known for nearly a hundred years that living things obey scaling relationships. Max Kleiber first recognized that the metabolic rates of different mammals scale according to their masses raised to a $3/4$-power \cite{ARB:Kle32}. More recently, Geoffrey West and his colleagues have provided a theoretical explanation for this scaling law, as well as for many other allometric laws found in biology \cite{ARB:Wes97,ARB:Whi2006}. Their theory is based upon the fractal branching networks (such as circulatory systems) found in all living things, whose function is to convey energy and nutrients to all parts of the organism.  They argue that the larger the organism, the more efficient the system that can be constructed to provide energy, thereby yielding the observed sublinear exponent of $3/4$.

More recently, West and his team examined a variety of properties of cities. They found that cities, which have long been compared to living things \cite{ARB:Zipf49,ARB:Jacobs01,ARB:Aristotle98}, obey scaling relationships as well \cite{ARB:Bet2007}. Similar to living things, cities have economics of scale, yielding sublinear scaling for such quantities as the number of gas stations within a city as a function of its population. In other words, you need fewer gas stations per person, in a bigger city. Examples of such scaling laws are shown in the upper portion of table \ref{table_1}. 

On the other hand, cities also exhibit superlinear scaling, which appears in relation to sociological quantities. As shown in the lower part of table \ref{table_1}, properties of cities related to economic productivity and creative output have exponents that are all found to cluster between 1 and 1.5, with the mean around 1.2. Thus, the productivity \textit{per person} increases as a city gets larger. However, this superlinear scaling has not been given a satisfactory mathematical explanation.

\begin{table}
\caption{A variety of urban quantities and their exponents. For instance, if $y$ denotes the number of gas stations in a city of population $N$, the data show $y=c N^{\alpha}$, with $\alpha \approx 0.77$. Table from Bettencourt et al \cite{ARB:Bet2007}.}
\centering
\begin{tabular}{l d}
Urban Indicators ($y$) & \text{Exponent } (\alpha) \\
	\hline
Gasoline stations & 0.77 \\ 
Gasoline sales & 0.79 \\ 
Length of electrical cables & 0.87 \\ 
Road surface & 0.83 \\
	\hline 
New patents & 1.27 \\ 
Inventors & 1.25 \\ 
Private R \& D employment & 1.34 \\ 
ÔSupercreativeÕ employment & 1.15 \\ 
R \& D establishments & 1.19 \\ 
R\& D employment & 1.26 \\ 
Total wages & 1.12 \\ 
Total bank deposits & 1.08 \\ 
GDP & 1.15 \\
\end{tabular}
\label{table_1}
\end{table}

Here we suggest a theoretical explanation for the superlinear
relationship between population size and innovation found in cities,
with a reasonable range for the exponent. Due to the sociological
nature of the variables being measured, it is natural to use a network
model of a city, since it is reasonable to assume that network effects
must underlie the superlinear scaling, as West and his colleagues have suggested \cite{ARB:Bet2007, ARB:Bet2008}.
For this, we draw on a recent class of models that derives 
superlinear scaling and {\em densification properties} 
from hierarchically organized networks \cite{ARB:Les2005},
adapting these models to the present question of productivity.

\section{Model and Results}

We first assume that all social interactions and relationships are
arranged in a hierarchical tree structure
\cite{ARB:Kle2001,ARB:Les2005,ARB:Wat2002}. 
Picture a binary tree, or in general,
a tree where each branch splits into $b$ new branches. For example, in
a city, each person is in a household, and there are many households
on a block, and many blocks in a neighborhood, and so forth. Or the
grouping could be based on your family tree, or corporations, or many
other ways to group individuals. While in reality each individual
belongs to many independent hierarchies \cite{ARB:Wat2002},
here we simplify it as a
single hierarchy, with branching number $b \geq 2$.  
We define the {\em distance} $d$ between two individuals in this 
hierarchy to be the height of their lowest common ancestor.
We view the total system as a city, meaning that a city of
population $N$ represents a single tree that contains $N$ leaves. On
top of the tree structure, which serves to determine the social
distance among nodes, a random graph is placed showing 
the social connections --- who actually knows whom.

\begin{figure}
\includegraphics[width=3 in]{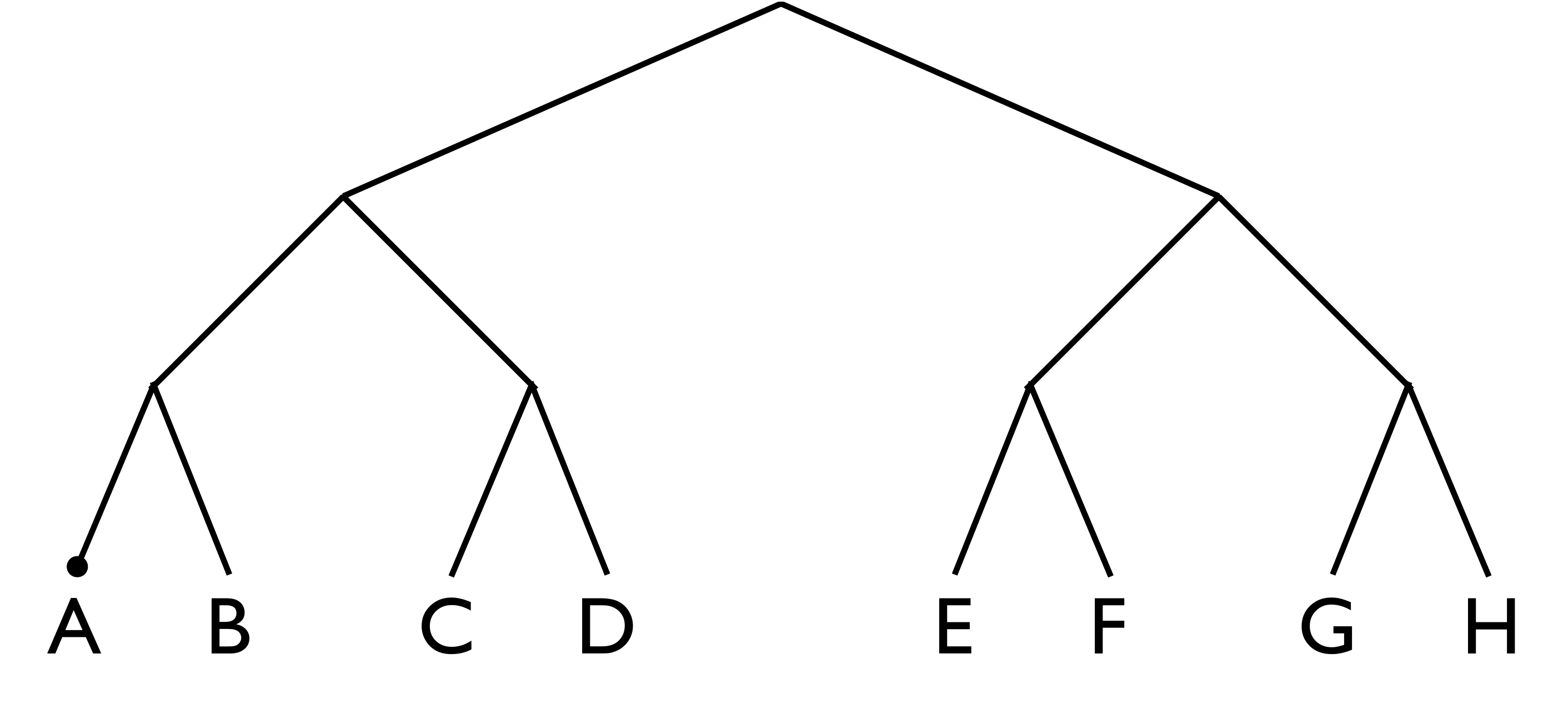}
\caption{\label{fig_1}Network representation of the social structure of a city's inhabitants. For example, if we are determining distance from individual A, then B is at distance 1, while G is at a distance of 3.}
\end{figure}

Our modeling strategy is to use these social connections as the basis
for a city's productivity: at a high level, 
we assume that each interaction
between a pair of people contributes to the overall productivity,
in a way that depends on the distance $d$ as 
measured within the hierarchy.
More concretely, our procedure for generating networks will
produce a directed graph, and we will account for the
productivity benefits of each edge $(v,w)$ by allocating
it to $v$'s overall productivity.  
This allocation to $v$
rather than to both nodes is essentially
for purposes of analysis, since we will be focusing
in the end on the total population's productivity rather than
any one individual's, and for determining this total we will
see that it does not matter to which individual we allocate
the benefits of the edge $(v,w)$.

The total creative productivity of the city is defined
to be the sum of the productivities of each individual,
and so we first consider how to compute individual productivities.
To calculate the total productivity of a single person, three separate
effects must be considered: (1) the probability of connecting to
an individual at distance $d$; (2) the number of available people at
distance $d$; and (3) the creative output that is obtained by linking to a
single person at distance $d$. Multiplying these together gives the
productivity due to one person linking to all of his collaborators at
distance $d$, as seen below: 
\begin{equation} \left [ \dfrac{\text{\#
contacts at }d}{\text{\# people at }d} \right ] \left [ \text{\#
people at }d \right ] \left [ \dfrac{\text{output at
}d}{\text{contacts at }d} \right ] \end{equation} 
By summing this term
over all distances, the total creative contribution of a single
individual is obtained. The functional form of each term in the above
recipe for calculating the productivity of a single individual is
discussed below.

Taking the first term, the social connections between collaborators
are constructed such that the likelihood of forming a connection at a
certain social distance drops off exponentially fast with distance
\cite{ARB:Kle2001,ARB:Les2005,ARB:Wat2002}.
That is, the probability of a connection being made between nodes of a
social distance $d$ (where $d$ is the height of the first common
internal node) is assumed proportional to $b^{-\alpha d}$, where
$\alpha$ is a tunable parameter greater than or equal to zero.

It is natural that the connection probability should decay with social
distance, but why exponentially? We have assumed that the social
network tree is self-similar at all levels (values of $d$). Since the
tree is self-similar, it makes sense to have the function also be
self-similar (scale-free) with respect to the value of $d$, and doing
this yields an exponential function (this assumption is relaxed in the
next section).

Since at each increase in $d$ there are exponentially more potential
contacts to interact with, we multiply the above function by a second
term, $b^{d}$, which means that as we increase $d$, while the
likelihood of making a connection decays, there are exponentially more
contacts to make. To keep things simple, we suppose connections are
only made between residents of the city (connections outside a city
are viewed as contributing less directly to the city's total productivity, 
and are ignored).

Lastly, the usefulness of a social connection within a city is assumed
to vary with its social distance. For example, one could assume that there
is a productivity benefit as social distance increases. 
This can be explained as
being due to the fact that individuals that are socially distant are
exposed to different ideas and experiences, and that collaboration
between two more socially distant individuals is more productive than
interaction between ones that are closer. However, the value of a
social connection is left open, and simply assumed to be proportional
to $b^{\beta d}$, where $\beta$ is a tunable parameter that can hold
any value (even negative values, allowing the value of a connection to
decrease with distance). An exponential function is reasonable here as
well, if we assume that a connection's innovation potential depends
on the number of individuals that lie between the two endpoints
of the connection in social space.
This assumption is also relaxed in the next section.

The total productivity of the social connections within
an $N$-person city is now a random variable equal to the
sum of all the individual productivities.
Its expectation $P(N)$ is given by 
\begin{equation}
P(N)=N \sum_{d=1}^{\log_b N} b^{-\alpha d}b^d b^{\beta d}
\end{equation}
In summary, the first term, $b^{-\alpha d}$, is the probability of
connecting at distance $d$. The second term, $b^{d}$, is the number of
nodes at distance $d$. And the final term, $b^{\beta d}$, is
productivity per connection. So, when these are multiplied together
and then summed for each distance, they yield the expected productivity
of one node in the full network. When multiplied by $N$ we get the
productivity for the entire network.

\begin{figure}
\includegraphics[width=3 in]{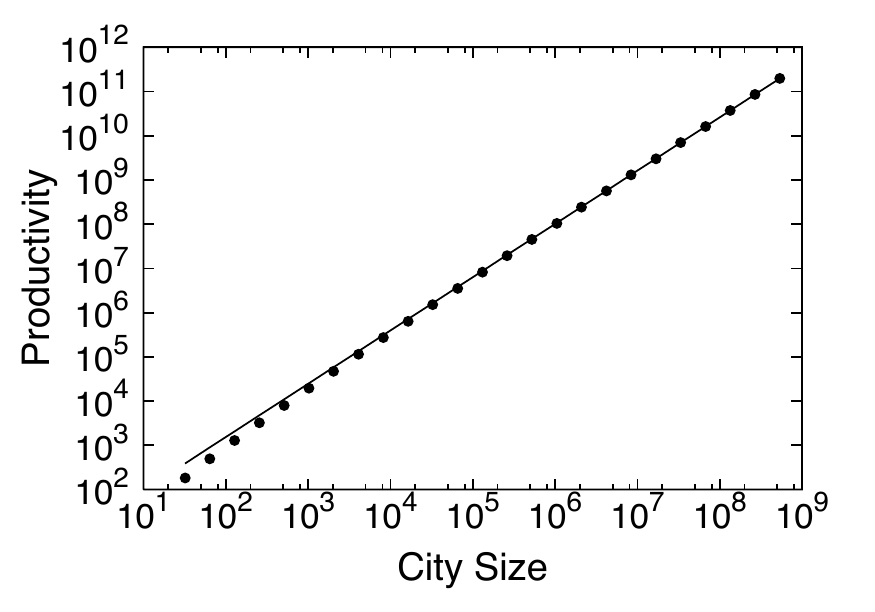}
\caption{\label{fig_2}Simulation and fit for $P(N)$. The points show the value of $P(N)$, calculated for  $\beta = 0.3$ and $\alpha = 1.1$. The least-squares approximation of the exponent is 1.205, and the expected value is $\beta - \alpha + 2 = 1.2$, for large values of $N$.}
\end{figure}

This can be summed exactly since it is a finite geometric series, and we get the following solution:
\begin{equation}
P(N) = N \left ( \dfrac{b^{\beta + 1}}{b^{\beta + 1} - b^{\alpha}} N^{\beta - \alpha  + 1} - 1 \right )
\end{equation}
For large values of $N$, we find $P(N)$ is proportional to
$N^{\beta-\alpha+2}$, if $\alpha < 1 + \beta$.
On the other hand, if $\alpha > 1 + \beta$, the function $P(N)$ becomes linear,
because the geometric series converges to a constant as $N$ becomes
large: \begin{equation}
P(N)=N \sum_{d=1}^{\log_b N} {(b^{\beta - \alpha + 1})^d} \xrightarrow[N \gg 1] {} N \dfrac{b^{\beta - \alpha + 1}}{1 - b^{\beta - \alpha + 1}}
\end{equation}
This is found to be in good agreement with numerical evaluation of the above summation, as can be seen in Figure \ref{fig_2}.

The growth of the productivity function with distance is not essential. 
The key is, rather, that as a city increases in size, it is
more likely to contain socially distant contacts. 
Having $\alpha < 1$ means that the expected number of contacts
scales superlinearly \cite{ARB:Les2005}.
Correspondingly, even if $\beta$ is slightly negative 
(meaning that more distant connections
are less productive), as long as $\beta - \alpha + 1 > 0$,
the densification of the social network due to the increasing
size of the city means that the productivity will grow superlinearly.
For the special case when $\beta$ is zero (all connections are equally
beneficial), the exponent $\beta - \alpha + 2$ must lie strictly
between 1 and 2, which is where all the measured exponents for urban
innovation lie. This is because we are assuming that $0 < \alpha < 1 +
\beta$ to get superlinear behavior, which means that $\alpha$ is
between 0 and 1.

Of course, other parameter relationships are also capable of yielding
the expected range of superlinear exponents.
For $\beta > 0$,
there is an exponentially decreasing probability of connecting to
someone at a social distance $d$ away, but connections at this
distance confer a productivity benefit that is 
exponentially increasing in $d$. 
In general, this model is a reasonable explanation
for the values observed within cities related to productivity and
innovation, and can be fit properly to explain the superlinear
exponents observed within cities.

\section{Expansion of the Analysis}

The assumptions of exponentials for the three functions that make up
the sum discussed above are stringent ones. What happens if we relax
these assumptions?

\begin{figure}
\includegraphics[width=3 in]{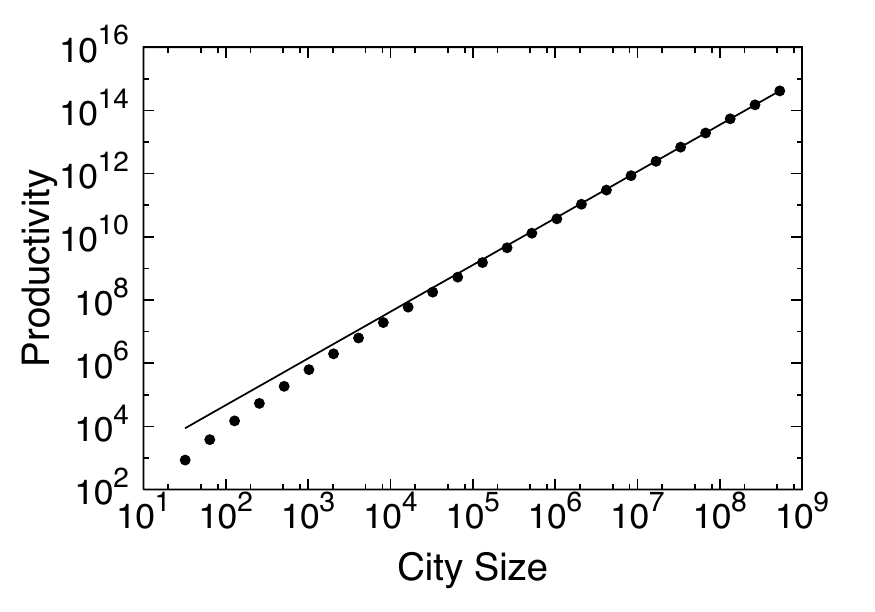}
\caption{\label{fig_3}A 'linearithmic' function. Using the function $b^{-\alpha d} b^{d} (d \ln{d})$  as the term within the sum for $P(N)$ (where $\alpha = 0.6$), the resulting function mimics a power law, with an exponent of 1.48.}
\end{figure}

Using a numerical simulation, each of the components of the sum can be modified, and we can graph the resulting scaling relationship and see if it remains superlinear. And in fact, the model is robust under a variety of situations. For example, instead of using an exponential for the creative benefit function, if we use a ÔlinearithmicÕ function ($d \ln{d}$), the resulting function asymptotically approaches a superlinear function, as seen in Figure \ref{fig_3}. A similar superlinear result can be obtained by replacing the function for the number of nodes at distance $d$ with a linearithmic function and leaving the other two functions exponential (by doing this, we are implicitly changing the structure of the social distance tree, such that the number of nodes no longer grows exponentially with distance).

\begin{figure}
\includegraphics[width=3 in]{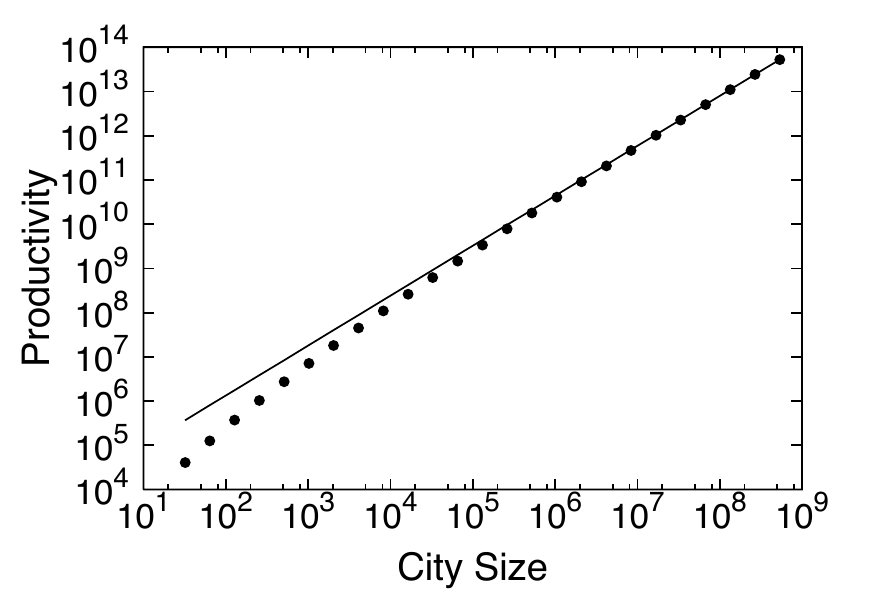}
\caption{\label{fig_4}Linear Functions. Using the function $(50 - \alpha d) (1 + \beta d) (2 d)$ as the term within the sum for $P(N)$ (where $\alpha = 0.4$ and $\beta = 0.1$), the resulting function still mimics a power law, with an exponent of 1.13.}
\end{figure}

In fact, even if all three functions are linear, the sum still grows
superlinearly with $N$, as seen in Figure \ref{fig_4}. Indeed, if the 
function is proportional to $d^3$, using the Euler-MacLaurin
summation, we find that $P(N) \approx N (\log_b{N})^4$, which grows a
bit faster than linearly.

\begin{figure}
\includegraphics[width=3 in]{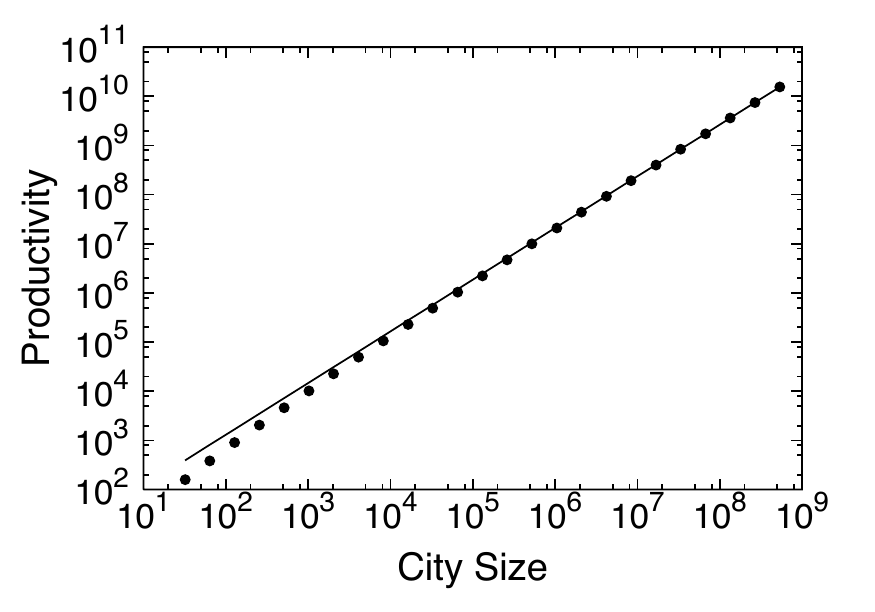}
\caption{\label{fig_5}The function $P(N) = N \log_b{N}$ mimics a power law, with an exponent of 1.05.}
\end{figure}

However, if the average productivity per node grows with $N$ but only at the rate $\log{N}$, then the rate of growth is only slightly superlinear, mimicking a power law exponent of 1.05, as seen in Figure \ref{fig_5}. Slightly faster than logarithmic growth for the summation appears to be required for superlinear growth of $P(N)$.

What can be seen is that using fairly loose assumptions, superlinear
growth can be obtained. Notably, these functions need not be power
laws. They can simply be superlinear functions (such as $P(N) \approx
N (\log_b{N})^x$), that mimic power laws. 
It is possible that this could be true 
of the observed city productivity data as well --- that is, 
it is possible that the productivity
functions observed are superlinear, but not necessarily power
laws.  Further measurement could help resolve this question.

\section{Discussion}

Ultimately, the heart of the model is the relationship between
long-distance ties and productivity in large cities.
These long-distance ties, which are
prevalent in a higher proportion when there is a larger population,
provide the potential for productive social interactions.

Granovetter's classic paper `The Strength of Weak Ties' considers this
explicitly \cite{ARB:Gra73}. As part of his study, Granovetter
examined the structure of Boston's West End and its inability to
organize against a neighborhood urban renewal project, which included
the large-scale destruction of buildings to make room for new
residential high-rises \cite{ARB:Col2005}. While the West End
contained many strong ties, since most individuals had been
lifelong residents of the area,
these strong ties often resulted in cliques,
where everyone was connected within a single group. Crucially,
however, Granovetter argues that there were few, if any, ties {\em between}
these tightly-knit local cliques. Since personal ties are generally necessary
for information spread and organizational ability (or as Granovetter put it, ``people rarely act on mass-media information unless it is also transmitted through
personal ties,'' \cite{ARB:Gra73}, p. 1374), the inhabitants of the West
End would have had a great deal of difficulty in organizing their
opposition to the municipal project. In contrast, if there had been
interaction throughout the social hierarchy, such as between
communities within the neighborhood, the outcome might have
been much different.
Along these lines, Charlestown, a similar Boston neighborhood, was
able to successfully organize against urban renewal. Granovetter
argues that there was a rich interconnection between different
communities, allowing for wider coordination. Alexander has similarly argued that rich interconnectivity between communities creates better cities \cite{ARB:Alex65}.

Of course, any good model must be testable in order for it to rise
above the level of a pleasant story. Additional work by the first author
\cite{ARB:Arb2008} has indirectly attempted to determine what the
value of $\beta$ is (it seems to be close to zero). Beyond this, given a
network of social interaction for a city, its hierarchical social
structure could be determined \cite{ARB:Cla2008}, to see if it conforms
to the type of growth with distance that is discussed above. This has
not yet been done, but it should be feasible, given the relevant datasets. Furthermore, it would be interesting to consider models as well as empirical data that consider interactions at scales larger than within a single city, such as between cities or within entire geographical regions.

In addition though, there are other possible explanations for this superlinear scaling in cities. For example, it could be that larger cities have a larger proportion of more highly educated individuals, which is enough to yield increased productivity per capita. Or it could be that larger cities simply have a greater transient population, which provides more fodder for different ways of thinking about the world, yielding a higher rate of productivity per individual. By distinguishing between our model based on social interaction and other competing models, we can get a better sense of how good our model is. But how can this be done?

While cities do exhibit superlinear scaling for a variety of quantities, many cities do not lie exactly on the predicted curve for a given property, based on a curve of best fit. For example, some cities will produce more patents than expected, while others will produce far fewer than expected, given their population. By looking at the pattern of social interaction in the underperforming cities as compared to the overperforming cities, we can determine how reasonable our model is. And by examining how well other models can predict this type of variation, as opposed to ours, we can determine what is the likeliest explanation for superlinear scaling within cities.

Nonetheless, as argued above, the presence of socially distant ties within a single city can be a powerful force. By using simple assumptions about social interactions, we gain a useful tool in understanding the mathematical behavior of innovation and productivity in cities.

\section{Acknowledgments}

We thank Geoffrey West, Luis Bettencourt, Stephen Ellner, and Adam Siepel for helpful discussions. Research supported in part by National Science Foundation grant DMS-0412757 to S.H.S.

\bibliography{cityscaling}

\end{document}